\documentclass[a4paper,12pt]{article}
 
\usepackage[sc]{mathpazo}
\usepackage{chicago}
\usepackage{pifont}
\usepackage{graphicx}
\usepackage{endnotes}
\usepackage{authblk}                                           
\usepackage{amssymb}
\usepackage{bbm}    
\usepackage{pgf}
\usepackage{tikz}
\usetikzlibrary{positioning,shapes,arrows,automata}
\usepackage{tensor}                                            
\usepackage{amsmath}

\usepackage{setspace}                           
\usepackage{hyperref}
\hypersetup{                 
    colorlinks=true,         
    linkcolor=blue,          
    citecolor=blue,           
    urlcolor=blue          
}
\usepackage{epigraph}
\usepackage[utf8x]{inputenc}
\usepackage[left=1in,right=1in,top=1in,bottom=1in]{geometry}                                  
\usepackage{enumitem}

\usepackage{multirow}

\let\oldmarginpar\marginpar
\renewcommand\marginpar[1]{\oldmarginpar{\color{red}\raggedright\scriptsize #1}}

\newcommand{\comment}[1]{}

\def\t{\ensuremath\mathrm}
\def\nn{\ensuremath\nonumber}

\newtheorem{theorem}{Theorem}

\newcommand{\qed}{\nobreak \ifvmode \relax \else
\ifdim\lastskip<1.5em \hskip-\lastskip
\hskip1.5em plus0em minus0.5em \fi \nobreak
\vrule height0.75em width0.5em depth0.25em\fi}

\title{{\bf \Large{Hawking Radiation and Analogue Experiments:\\ A Bayesian Analysis}}}

\author[1]{ \bf Radin Dardashti\thanks{email: \href{mailto:Dardashti@uni-wuppertal.de}{Dardashti@uni-wuppertal.de}}}
\author[2]{\bf Stephan Hartmann\thanks{email: \href{mailto:S.Hartmann@lmu.de}{S.Hartmann@lmu.de}}}
\author[3]{\bf Karim Th\'ebault\thanks{email: \href{mailto:karim.thebault@bristol.ac.uk}{karim.thebault@bristol.ac.uk}}}
\author[4]{\bf Eric Winsberg \thanks{email: \href{mailto:winsberg@usf.edu}{winsberg@usf.edu}}}

\affil[1]{\small{{\it Interdisziplin\"{a}res Zentrum f\"{u}r Wissenschafts- und Technikforschung/ Philosophisches Seminar}, Bergische Universit\"{a}t Wuppertal}} 
\affil[2]{\small{{\it Munich Center for Mathematical Philosophy}, Ludwig Maximilians Universit\"{a}t Munich}} 
\affil[3]{\small{{\it Department of Philosophy}, University of Bristol}} 
\affil[4]{\small{{\it Department of Philosophy}, University of South Florida}}  

\begin{document}

\maketitle

\begin{abstract} 
We present a Bayesian analysis of the epistemology of analogue experiments with particular reference to Hawking radiation. First, we prove that such experiments can be  confirmatory in Bayesian terms based upon appeal to `universality arguments'. Second, we provide a formal model for the scaling behaviour of the confirmation measure for multiple distinct realisations of the analogue system and isolate a generic saturation feature. Finally, we demonstrate that different potential analogue realisations could provide different levels of confirmation.  Our results  provide a basis both to formalise the epistemic value of analogue experiments that have been conducted and to advise scientists as to the respective epistemic value of future analogue experiments.   
\end{abstract}

\newpage

\tableofcontents
 
\section{Introduction}

Recent years have seen a proliferation of experiments designed to probe the phenomenon of Hawking radiation via analogue black hole systems. Reports on these experiments include claims of observation of classical, thermal aspects of Hawking radiation in an analogue white hole created using surface water waves \shortcite{weinfurtner:2011,weinfurtner:2013} and observation of quantum entanglement across an analogue horizon, as is expected from Hawking radiation, using a Bose-Einstein Condensate (BEC) \cite{steinhauer:2016}. Despite these successes, an uncontested and repeated experimental demonstration of the full analogue Hawking phenomenon is still to be provided.\footnote{For more on surface water wave experiments see \shortcite{rousseaux:2008,rousseaux:2010,michel:2014,unruh:2014,euve:2016,torres:2017,Euve:2018}. For further results and discussion of the Steinhauer's BEC experiments see \shortcite{steinhauer:2014,steinhauer:2015,Finke:2016,steinhauer:2016a,Nova:2018,Leonhardt:2018}.} The not fully settled experimental status of analogue Hawking radiation not withstanding, there is a clear need for methodological appraisal of what, in principle, can be established about black hole Hawking radiation based upon  analogue experiments. The most  pressing questions is whether or not there are circumstances in which analogue experiments can be taken to provide inductive support for conclusions about astrophysical black holes. Can analogue experiments provide  `experimental confirmation of Hawking's prediction' (Jeff Steinhauer quoted in Haaterz \cite{Efrati:2016}), or are they simply `amusing feat[s] of engineering' that `won't teach us anything about black holes' (Daniel Harlow quoted in {\em Quanta} \cite{Wolchover:2016}). 

In this paper we will substantially extend previous philosophical work characterising analogue black hole experiments as a form of `analogue simulation' \shortcite{dardashti2015confirmation,Thebault:2016} via application of Bayesian confirmation theory. In that previous analysis, emphasis was placed upon the \textit{qualitative} claim that certain `universality arguments' can be used to link evidence about the `source' dumb hole system to the `target' black hole system. The results of this paper are \textit{quantitive} in nature and licence normatively valuable conclusions regarding the structure of such inferences. 

We first present a theorem that demonstrates how the confirmation claim can be qualitatively characterised in Bayesian terms. The role of the universality arguments is understood in terms of support for a background assumption that is common between the source and target models. This means that there is a binary variable that can be assumed to be positively correlated with the empirical adequacy of both the source and target models. Evidence in favour of the model of the source system can thus be used to make inferences about the target system. Although not in-and-of-itself a validation of the argument for confirmation via analogue simulation, the Bayesian analysis demonstrates the internal consistency of the informal arguments given in \shortcite{dardashti2015confirmation}. Furthermore, the formal model provides a powerful evaluative and heuristic tool for the further analysis of the structure of the inferences involved in cases of analogue simulation.  In particular, in the remains of the paper, we present two further results that we expect will be of interest to contemporary analogue black hole experimentalists.   

Our second principal result relates to the behaviour of the confirmation measure in the context of experimental realisations of the analogue system in different types of media. This is of particular relevance to contemporary analogue black hole research where a diverse range of analogue realisations of Hawking radiation are being pursued: e.g., surface waterwaves, BECs, superfluid helium-3, moving optical media. The immediate question in this context is how the number of distinct types of analogue system one constructs relates to the confidence one should have in the astrophysical effect. The second key result of this paper is a formal model for `multiple source' analogue simulation displaying the generic feature of `saturation' in confirmatory power with an increase in the number of sources. Significantly, the saturation in confirmation  indicates that, under plausible assignments of priors, even an extraordinarily large range of diverse analogue experiments will not lead to conclusive confirmation of astrophysical Hawking radiation. This is in tune with the scientific intuition that there is a limit to what can be learned about astrophysical Hawking radiation via analogue experiments.  

Finally, and perhaps most intriguingly, there is the question of whether different potential analogue realisations could provide different levels of confirmation. Would we learn more about astrophysical black holes from an analogue experiment based upon liquid helium or BECs? The third key result of the paper is a theorem proving that analogue experiments in which we are more confident about the fundamental physics (e.g. BECs) teach us less about the target system than those about which we are less confident (e.g. superfluid helium-3). Our results thus provide a basis to both formalise the epistemic value of analogue experiments that have been conducted and to advise scientists as to the respective epistemic value of future analogue experiments. As such, our work demonstrates the enduring value of the Bayesian framework as a tool for analysing the protean forms of scientific inference.     

\section{Confirmation, Analogy and Experiment}

The literature on analogical reasoning in science is fairly extensive, with particularly noteworthy contributions by Keynes \citeyear{keynes:1921}, Hesse \citeyear{hesse:1963,hesse:1964,hesse:1974}, \citeN{bailer:2009} and Bartha \citeyear{bartha:2010,bartha:2013}.\footnote{See also \citeN{norton2011analogy} for an importantly different take on analogical arguments. Norton's analysis focuses on analogical arguments that proceed via subsumption of the target system into a larger class of entities, including the source system. There are broad parallels between the structure of such inferences and our analysis.} Typically analogical arguments have the form of abstract speculative inferences regarding possible features of one system (`the target') based on known features of another system (`the source'). Classic examples are Reid's argument for the existence of life on other planets based upon life on earth \shortcite{reid:1850} or Hume's argument for animal consciousness based upon human consciousness  \shortcite{hume:1738}.  

Analogical arguments evidently play an important \textit{heuristic role} in scientific practice in that they provide `cognitive strategies for creative discovery' \shortcite[p. 56]{bailer:2009}. The \textit{epistemic role} of analogical arguments in science is, however, more controversial. In particular, the literature contains diverging answers regarding whether analogical arguments can provide Bayesian confirmation of a hypothesis regarding the target system. From a Bayesian perspective on confirmation, evidence for a hypothesis can count as confirmatory only if the probability of the hypothesis given the evidence together with certain background assumptions is larger than the probability of the hypothesis given only the background assumptions. In a detailed and nuanced treatment of the issue, \citeN[pp. 208-19]{hesse:1974} suggests that analogical arguments \textit{can in some cases} be confirmatory in a Bayesian sense, so long as the analogical relationship that holds is in terms of what she calls a `material analogy'. That is, where there is a similarity relation of sharing at least one predicate between the target and source systems.\footnote{\citeN[p. 216]{hesse:1974} explicitly rules out the possibility of confirmation obtaining in cases where there is purely a `formal analogy'. That is, where target and source are both interpretations of the same formal calculus but do not share material similarities.} 

Contrastingly, Bartha \citeyear{bartha:2010,bartha:2013} offers arguments that analogical arguments \textit{cannot in principle} be confirmatory in a Bayesian sense. In particular, he suggest that because the information encapsulated in an analogical argument is taken to already be part of the `background knowledge', the probability of a hypothesis regarding the target system must be identical before and after including the analogical argument. Bartha takes this instance of the familiar problem of old evidence \cite{glymour:1980} to be significant enough to bar analogical arguments from being confirmatory in Bayesian terms.  Rather, following Salmon \citeyear{Salmon:1967,Salmon:1990}, Bartha argues that arguments by analogy can establish only the \textit{plausibility} of a conclusion in the technical sense of justifying the assignment of a non-negligible prior probability assignment \cite[\S8.5]{bartha:2010}.
 On this analysis, it is not \textit{in principle} possible for analogical arguments to confer inductive support for a hypothesis. That is, although analogical arguments can certainly be stronger or weaker, even the strongest possible analogical argument cannot confer confirmation in a Bayesian sense: they are abstract inferences that can only ever support plausibility claims rather than providing inductive evidence.  

Although worthy and insightful, the treatments of \citeN{hesse:1974} and \citeN{bartha:2010} do not extend to the analysis of analogue experiments. This is because analogue experiments are unlike arguments by analogy in exactly the respects that are crucial for either the Hesse or the Bartha analysis to go through. The question of whether one agrees with Hesse or Bartha about the confirmatory power of  arguments by analogy is simply tangential to the analysis of the confirmatory power of analogue experiments. In the case of Hesse, this is indicated by the fact that her notion of material analogy is too strict to accommodate the subtle relation that the model of the target has to the model of the source in the case of analogue black hole experiments. We are dealing with an \textit{analogue simulation} that does not involve a material analogy in the sense of Hesse since there is not a \textit{physical property} common between the target and source systems. Rather than a material relation between systems, we have a syntactic isomorphism between models.\footnote{Here `syntactic isomorphism' is a natural generalisation of Hempel's \citeyear{hempel:1965} notion of nomic isomorphism to the case of a relation between models rather than laws.}$^,$\footnote{\citeN{hesse:1963}, does in fact, rather presciently, consider the relevance of simulators in her account of models and analogies in science. Tantalisingly, she says that analogue machines (i.e. simulators): `are useful and necessary as predictive models precisely in those cases where the material substance of parts of the analogue is \textit{not essential} to the model, but where the mutual relations of the parts are essential' (p. 102) This connection is unfortunately not taken up in the 1974 Bayesian analysis.} 

Bartha's negative analysis of the prospect for confirmation via analogical argument is similarly inapplicable to analogue experiments. This is because analogue experiments, unlike analogical arguments, are essentially empirical activities: they involve learning about the world by manipulating it. In analogue black hole experiments we manipulate the source such that certain explicit modelling assumptions matching those for the target obtain. Analogue simulation thus resembles a form of experimentation, involving the `programming' of a physical system such that it can be used to `simulate' another physical system. Thus, we see that conclusions from the philosophical analysis of analogical argument should not be taken to be readily extendible to cases of analogue simulation in contemporary science. In particular, it is self-evidently the case that the old evidence problem for the Bayesian analysis of traditional arguments by analogy \`a la Hesse, is not longer relevant for analogue experiments. Analogue experiments unlike analogical reasoning explicitly involve the collection of new evidence and there are not good grounds for relegating their significance to mere plausibility.

This collection of new evidence motivates us to consider the `epistemology of analogue experimentation' in parallel with the epistemology of conventional  experiments. As has been noted by various authors \cite{franklin:1989,winsberg:2010,Franklin:2016}) conventional experiments are generally only of epistemological significance when supplemented by arguments that imply that   the information learned about the system being manipulated (`the source') is relevantly probative about the class of systems that are of interest to the experimenters (`the target'). A nice illustration of this point is provided by experiments designed to learn about the thermal conductivity of the iron in Earth's core \shortcite{konopkova:2016,dobson:2016}. The experiments were carried out \textit{in the lab} using samples of iron that are placed in a laser-heated diamond-anvil cell. The pressure and temperature that the iron samples were subjected to were specifically matched to those relevant to the cores of Mercury-sized to Earth-sized planets. For the experiments to achieve their epistemic purpose, they must be supplemented with arguments that the measurements are be `relevantly probative' of the thermal conductivity of iron in the core's of Mercury-sized to Earth-sized planets.  That is, there must be a basis to generalise from the observations regarding the lab based system (the `source'), to relevant systems outside the lab (the `target'). In parallel, what we take to be the key question in the epistemology of analogue experimentation is  whether we can provide arguments that the relevant source systems `stand-in' for the target systems to which the analogical relationship refers. Can we find arguments that analogue black hole experiments  are relevantly probative of the thermal properties of astrophysical black holes?\footnote{See \cite{Thebault:2016} for an analysis of the connection between the epistemology of analogue experiments and the notions of internal' and external validation as discussed in the philosophy of experimentation.} 
 
\section{Hawking Radiation and Universality} \label{unschutz}

Hawking radiation \cite{hawking:1975} is a thermal phenomena that is predicted to be generically associated with black holes. Despite the absence of either a clear causal process behind the radiation or experimental evidence, it is widely believed in by theoretical physicists. In fact, the ability to recover Hawking radiation could even be taken as a theoretical test of prospective theories of quantum gravity, much like the recovery of the Bekenstein-Hawking formula for black hole entropy \cite{wuthrich:2017}. There are two connected reasons why physicists are so confident in the prediction of Hawking radiation. First, given the Unruh effect \cite{unruh:1976}, which associates a temperature with acceleration, Hawking radiation seems to be directly implied by the equivalence principle.\footnote{Such a conclusion is, in fact, a little too quick since the equivalence principle holds only locally and Hawking radiation is a global effect. See \cite{helfer:2010}.} Second, starting from Hawking's original calculation a remarkable number of different derivations of the effect have been given.\footnote{ See \cite{leonhardt:2008,thompson:2008,Wallace:2017a}.} The consensus is that the effect is `remarkably robust' to the addition of complicating factors to the original derivation. The overall implication is that very general theoretical constraints coming from quantum field theory and general relativity (two well tested theories) necessitate that something like Hawking radiation must exist. The purpose of this paper is not to address the evidential import of such theoretical considerations. Rather, our focus is on the potential for analogue experiments to provide confirmatory evidence of a form akin to conventional experiments. This notwithstanding, questions of robustness will return to the fore in the context of a particular form of universality argument that will be found to be central for questions of confirmation. Before then, it will be instructive to consider the basic elements of the original Hawking derivation of a radiative flux for astrophysical black holes in comparison with their sonic analogues. 

Hawking's analysis is performed in the context of a  semi-classical approach to gravity. That is, we consider matter as described by quantum field theory and spacetime as described by a continuous classical geometry. Crucially, although the spacetime   in question can have non-trivial curvature, it is not coupled to the quantum field. That is, there can be no `back-reaction' between the quantum matter and classical geometry. For this modelling framework to be valid it is assumed that we are considering quanta of wavelengths much larger than the Planck length. Quanta of the order of the Planck length could be expected to `see' the (presumed) non-classical and non-continuous structure of spacetime and would necessitate a quantum theory of gravity in their description. Quite general formal considerations can be used to show that in the semi-classical framework the vacuum state of a quantum scalar field defined at past null infinity need not appear as a vacuum state to observers at positive null infinity. In particular, it may contain a `particle flux'. What Hawking shows in his original paper is that for spacetime in which an astrophysical black hole forms there will be a particle flux which observers at positive null infinity will associate with the black hole horizon. The asymptotic form of the expression for the particles flux is shown to depend only upon the \textit{surface gravity} of the black hole denoted by $\kappa_{G}$. Surface gravity is essentially the force per unit mass that must be applied at infinity in order to hold a stationary zero angular momentum particle just outside the horizon \cite{Jacobson:1996}. Hawking's calculation implies that a black hole has intrinsic properties that are connected to a non-zero \textit{thermal} particle flux at late times. The precise relation takes the form:
\begin{equation}
\langle \hat{N}_\omega \rangle = \frac{1}{{\rm e}^{\frac{2 \,  \pi \, \omega}{\hbar \, \kappa_{G}}}-1} \quad  {\rm with} \quad T_{BH} := \hbar \,  \kappa_{G}/ 2\pi,
\end{equation}
where $\hat{N}_\omega$ is the number operator for modes detected at late times with frequency $\omega$ and $\hbar$ is Planck's constant divided by $2\pi$. 

One key feature of the derivation of the temperature is worth noting here since it will be very important in what follows. In the derivation of Hawking radiation an exponential gravitational red-shift means that the black hole radiation detected at late times must be taken to correspond to extremely high frequency radiation at the horizon. These `trans-Planckian' modes are of wavelengths that are of precisely the kind that we presumed to exclude in using the semi-classical framework. There is thus a tension between the initial modelling assumptions and the details of the calculation. We will return to this issue shortly.

Not long after the derivation of Hawking's radiation, it was proposed by Unruh that a similar thermal effect might exist in the context of sound in fluid systems \cite{unruh:1981}. In particular, Unruh showed that the key elements of Hawking's calculation could be re-applied in the context of a semi-classical model of sound in fluids. The role of the spacetime geometry is now played by a `bulk' fluid flow as described by continuum hydrodynamics. The role of the quantum field is then played by a quantized linear fluctuation within the fluid, a phonon. The modelling framework of continuum hydrodynamics is only valid provided fluid density fluctuations of the order of molecular lengths can be ignored. So for this semi-classical description to be adequate the wavelengths of the phonons must be much larger than the intermolecular distances. Unruh's brilliant insight was to recognise that there is a special class of analogue fluid systems for which the equations of semi-classical continuum hydrodynamics take a form isomorphic to those of semi-classical gravity.  The role of the black hole event horizon is now played by an effective acoustic horizon where the inward flowing magnitude of the radial velocity of the fluid exceeds the speed of sound.  The black hole is replaced by a \textit{dumb hole}. Just as in the gravitational Hawking effect a black hole event horizon is associated with a late time \textit{thermal photonic flux}, in the hydrodynamic Hawking effect Unruh showed that a dumb hole sonic horizon can be associated with a late time \textit{thermal phononic flux}. 

An alternative medium for constructing acoustic black holes, that behaves exactly like a fluid in an appropriate limit, is give by a Bose-Einstein condensate \shortcite{garay:2000}. This is an exotic form of matter predicted in the 1920s \cite{bose:1924,einstein:1924} but not created experimentally until 1995 \shortcite{Anderson:1995}. Crucially, in the limit of \textit{weak coupling} and \textit{no backreaction}  density fluctuations in a BEC are described by the Gross-Pitaevskii equation. When variations in the density of the BEC occur on length scales much greater than the \textit{healing length}, the Gross-Pitaevskii equation can be used to derive an equation of the same form as that of an irrotational fluid derived by continuum hydrodynamics. Following the same line of reasoning as Unruh's original ideal fluid argument, \shortcite{garay:2000} derived a  BEC Hawking Effect using appeal to this hydrodynamic approximation to a BEC. 

There are now a huge number of potential analogue realisations of the Hawking effect: phonons in superfluid helium-3, `slow light' in moving media, traveling refractive index interfaces in nonlinear optical media, laser pulses in nonlinear dielectric medium.\footnote{See \shortcite{jacobson:1998,philbin:2008,belgiorno:2010,unruh:2012,liberati:2012,nguyen:2015,Jacquet:2018}.}  To realize the Hawking effect it seems it is sufficient to have: i) a classical (effective) background with quantum fields living on it; and ii) an (effective) geometry with an (effective) causal horizon. 

A crucial feature in the derivation of all these effects is the integration over extremely high energy `trans-Planckian' modes.  As noted above, in the black hole case these modes get included in the calculation due to an exponential redshift between the horizon (where they originate) and the late time, far distance limit (where they are detected). Such a feature is generic to all models of Hawking radiation in which the modes originate near the horizon.\footnote{It is worth noting here that whilst, the non-standard derivation of \cite{giddings:2016} does appear to allow one to avoid this feature, that of  \cite{polchinski:1995}, prima facie, does not \cite[pp. 37-8]{harlow:2016}.} Since the `trans-Planckian' regime beyond the domain of applicability of the semi-classical modelling framework we are using, this problem of exponential redshift seems rather worrying. In fact, according to some, the trans-Planckian problem is so serious as to cast doubt upon the Hawking calculation entirely.  Unruh, for instance, even asks that `if the derivation relies on such absurd physical assumptions, can the result be trusted?' \cite[p. 534]{unruh:2014}. The problem with `trans-Planckian' modes has a direct analogue in both the continuum hydrodynamic and BEC derivations. In particular, the acoustic analogue of the gravitational redshift, means that in both cases we are including in our calculation phonons of wavelengths small enough to probe the regimes very far beyond the inter-molecular length and healing length respectively. 

Fortunately, there are good reasons to expect that the Hawking effect in both gravitational and analogue cases will be robust to disturbance from trans-Planckian modes. In particular, \citeN{unruh:2005} have provided strong theoretical reasons to expect that, under certain conditions, any modifications to the Hawking flux by trans-Planckian modes will be negligible.\footnote{For further work on these issues, using a range of different methodologies, see for example \cite{corley:1998,himemoto:2000,barcelo:2009,coutant:2012}. For philosophical discussion see \cite{dardashti2015confirmation} and \cite{Gryb:2018}.} Unruh and Sch\"{u}tzhold show that a wide family of trans-Planckian effects can be factored into the calculation of Hawking radiation via a non-trivial dispersion relation. To lowest order and given certain modelling assumptions, Hawking radiation, both astrophysical and acoustic, is independent of the details of the underlying physics.  

A significant distinction that can be made in this context is between robustness and universality.\footnote{Here we are drawing upon the account of \shortcite{Gryb:2018} who in turn are largely following the discussion of \cite{batterman:2000}.}  Robustness is the insensitivity of a phenomenon under a \textit{token-level} variation with respect to different possible micro-physics in a single type of system. Universality is the insensitivity of a phenomenon under a \textit{type-level} variation between systems with fundamentally different material constitution (e.g. BECs and a classical fluid). Given these definitions, we can plausibly take the work of Unruh and Sch\"{u}tzhold to be an argument for \textit{both} the robustness and the universality of the Hawking effect. We can also, on this basis, draw a suggestive parallel to the notion of `universality' found in the context of condensed matter physics. That said, nothing in our analysis will hang upon the particular characterisation of universality.\footnote{\shortcite{Gryb:2018} investigates these parallels at length. Analogue experiments that might plausibly be described using our formal framework based upon Wilsonian universality arguments are \shortcite{thouless1989,prufer:2018,erne:2018,eigen:2018}.} Rather, inferentially speaking, what is important is that the `universality argument' in question provides support for a background assumption, common between the different models, that has moderate prior probability. The common background assumption supported by the argument of Unruh and Sch\"{u}tzhold is that in general ultra high frequency physics does not significantly effect the Hawking spectrum. We take it that the  Unruh and Sch\"{u}tzhold arguments imply that such a common background assumption should not have a very low prior, and the trans-Planckian problem implies that such a common background assumption should not have a very high prior. In general terms, there are good reasons to expect that universality arguments will perform something like this function, but nothing in what follows will depend upon this assumption.

\section{Bayesian Analysis}

\subsection{Single Source Confirmation}

The key claim that we wish to investigate is whether analogue `dumb hole' experiments can provide inductive support for hypotheses regarding black holes given we believe the appropriate universality arguments. In what follows we give a Bayesian network representation of the proposed inferential structure of analogue simulation defended in  \shortcite{dardashti2015confirmation} and show that the evidence in the source system can provide confirmation of hypotheses regarding the target system in certain circumstances.\footnote{For models of confirmation in terms of the Bayesian framework see \shortcite{hartmann2010bayesian,bovens2003bayesian} or for the hypothetic-deductive framework see \shortcite{betz2013revamping}. Throughout this paper, we follow the convention that propositional variables are printed in italic script, and that the instantiations of these variables are printed in roman script.} 

Let us start with the representation of the target system $\mathcal{T}$. We denote by $M$ a propositional variable that takes the two values:
\begin{description}
\item[$ \t M$:] The modelling framework $\mathcal{M}$ provides an empirically adequate description of the target system $\mathcal{T}$ within a certain domain of conditions $D_\mathcal{M}$.
\item[$\neg \t M$:] The modelling framework $\mathcal{M}$ does not provide an empirically adequate description of the target system $\mathcal{T}$ within a certain domain of conditions $D_\mathcal{M}$.
\end{description}
The adequacy of the modelling framework $\mathcal{M}$ depends on whether the background assumptions which justify the empirical adequacy of the modelling framework obtain. We denote with $X_M$ the random variable with the values:
\begin{description}
\item[$\t X_M$:] The background assumptions $x_M=\{x_M^1, x_M^2,\dots, x_M^n\}$ are satisfied for target system $\mathcal{T}$.
\item[$\neg \t X_M$:]  The background assumptions $x_M=\{x_M^1, x_M^2,\dots, x_M^n\}$ are not satisfied for target system $\mathcal{T}$,
\end{description}
where the second statement should be read in terms of a negation of the disjuncts. The role of the background assumptions is to define and justify the domain of conditions for the model. These assumptions involve knowledge, both theoretical and empirical, which goes beyond what is encoded within the model. Such knowledge need not be in the form of a simple, unified framework. Rather the background knowledge of the people who build and use models can contain an incompletely integrated set of explicit and tacit ideas about  when a particular modelling framework will be adequate for a particular purpose and to a particular desired degree of accuracy. 

With this in mind, we can introduce the random variables $A$ and $X_A$ for the source system $\mathcal{S}$ described by the modelling framework $\mathcal{A}$. The variable $A$ is a propositional variable that takes the two values:
\begin{description}
\item[$\t A$:] The modelling framework $\mathcal{A}$ provides an empirically adequate description of the source system $\mathcal{S}$ within a certain domain of conditions $D_\mathcal{S}$.
\item[$\neg \t A$:] The modelling framework $\mathcal{A}$ does not provide an empirically adequate description of the source system $\mathcal{S}$ within a certain domain of conditions $D_A$.
\end{description}
and $X_A$ is the random variable with the values:
\begin{description}
\item[$\t X_A$:] The background assumptions $x_A=\{x_A^1, x_A^2,\dots, x_A^k\}$ are satisfied for source system $\mathcal{S}$.
\item[$\neg \t X_A$:]  The background assumptions $x_A=\{x_A^1, x_A^2,\dots, x_A^k\}$ are not satisfied for source system $\mathcal{S}$.
\end{description}

The systems $\mathcal{S}$ and $\mathcal{T}$ are assumed to represent different \textit{types} of systems and thus have fundamentally different material constitutions. This means that the background assumptions behind the models $\mathcal{M}$ and $\mathcal{A}$ can reasonably be assumed to be very different. For example, compare the assumption of a curved but fixed spacetime for a black hole with the assumption of a flat spacetime for an analogue black hole; or compare the assumption of a continuous non-back-reacting spacetime for the black hole with the assumption of a continuous non-back-reacting bulk fluid flow for the analogue black hole. Given this, it is justified, prima facie, to assume that $X_M$ and $X_A$ are probabilistically independent (we will discuss the status of this assumption further shortly). Furthermore, we have assumed that the source system is empirically accessible meaning we can gain empirical evidence regarding (at least) some of its consequences. We can encode this by introducing a variable $E$  corresponding to the two values, $\t E$, the empirical evidence obtains, and $\neg \t E $, the empirical evidence does not obtain. 

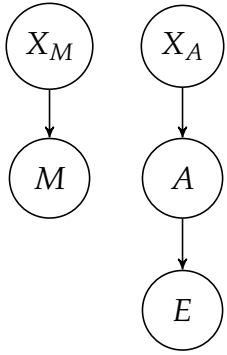
\begin{figure}
\centering
\scalebox{1}{\begin{tikzpicture}[->,>=stealth', auto, node distance=1.75cm, semithick]
\tikzstyle{every state}=[text=black]
	\node[state](Xm) {$X_M$};
	\node[state](Xa) [right of=Xm]{$X_A$};
	\node[state](M) [below of=Xm]{$ M$};
	\node[state](A) [below of=Xa]{$ A$};
	\node[state](E) [below of=A]{$ E$};
		
	\path[] (Xm) edge         node{}(M);
	\path[] (Xa) edge         node{}(A);
	\path[] (A) edge         node{}(E);
\end{tikzpicture}}
\caption{The one-source system without universality.} \label{fig2}
\end{figure}

We can represent all the variables introduced thus far as well as the probabilistic dependencies using a Bayesian network \cite{bovens2003bayesian}. The random variables are represented as `nodes' in the network (i.e. circles) and the probabilistic dependences as directed edges (i.e. arrows). We draw an arrow between two nodes when the variable in the `parent node' has a direct influence on the variable in the `child node'. Unconditional probabilistic independence is represented implicitly by the absence of an arrow between two nodes. The entire set up thus far is represented by the Bayesian network in Figure~\ref{fig2}. 

Consider the case of an analogue black hole experiment, leading an agent to learn $\t E$, in a context where the agent does not fully believe in any form of universality argument. Clearly, due to the probabilistic independence of $X_M$ and $X_A$ and of $M$ and $A$, we expect the independence of $M$ and $E$ and thus have that $P(\t M|\t E) = P(\t M)$. We therefore do not have confirmation. This is despite the fact that the equations of the source model are syntactically isomorphic to the equations that are supposed to govern the behaviour of the target model. The isomorphism is somewhat surprising, given that one arrives at the equations starting from very different areas of physics (i.e. semi-classical gravity and semi-classical continuum hydrodynamics) and by making different background assumptions. However, the agent has no reason to believe that the isomorphism represents a deeper fact about nature: on its own, the syntactic relation between the models gives us no reason to doubt the probabilistic independence of $X_M$ and $X_A$. 

There will, of course, always be available \textit{some} basis to connect certain of the background assumptions. For instance, common features such as the dimensionality of spacetime or the masslessness of relevant force carrying particles.\footnote{See \cite[\S3.2]{dardashti2015confirmation} for discussions of such common background assumptions in the context of $1/r^{2}$ force laws.} Thus, strictly speaking, background assumptions regarding different systems are never probabilistically independent. The reason why it is safe to neglect such common background assumptions for the purposes of confirmation is that they will generally be generic statements about which we have high confidence. This means that they cannot lead to any significant confirmation being transferred between $E$ and $M$ and can thus be neglected (see \cite[\S4.c]{HowsonUrbach:2006}).

Now, consider a context where the agent \textit{does} believe in a universality argument for Hawking radiation, such as that provided by Unruh and Sch\"{u}tzhold. The universality arguments give a physical justification for interpreting the syntactic isomorphism as indicative of a deeper fact about nature. In particular, it implies that we should assign a moderate prior to the common background assumption that ultra high frequency physics does not significantly effect the Hawking spectrum. It is crucial here that without the universality argument none of the \textit{type-specific} forms of this background assumption already have high priors. Consider a situation where without the universality argument we already had strong independent grounds on which to assign high probability to the background assumption that for a particular type of analogue system the ultra high frequency physics does not significantly effect the Hawking spectrum. This would mean that the relevant member of $X_A$, corresponding to the specific statement of independence from ultra high frequency physics for that system, would have very high prior. In such a case, not much inductive support would be added to the generic form of the statement by learning $\t E$. This would block the transfer of confirmation from $E$ to $M$. It is thus particularly significant that due to the generic nature of the trans-Planckian problem, the member of $X_A$ corresponding to the specific statement of independence from ultra high frequency physics will for every type of system have a relatively low prior. There is thus a clear route for strong inductive support to flow through the generic statement of independence from ultra high frequency physics.

With these considerations in mind, let us then introduce a new binary variable $X$ that has the values:
\begin{description}
\item[$\t X$:] Ultra high frequency physics does not significantly effect the Hawking spectrum.
\item[$\neg \t X$:] Ultra high frequency physics does significantly effect the Hawking spectrum.
\end{description}
$X$ expresses a rather general claim, which can plausibly be assumed to have neither very high nor very low credence (if we were certain about $X$, the inference from  $\mathcal{A}$ to $\mathcal{M}$ would be blocked. We will say more about this later). 

For the sake of simplicity, let us now subsume the background assumptions, that are not addressed by $X$ (or any other generic statement) under the relevant nodes. That is, we will subsume background assumptions exclusively concerning the target system $\mathcal{T}$ under $M$ and background assumptions exclusively concerning the source system $\mathcal{S}$ under $A$. 

\begin{figure}
\centering
\scalebox{1}{\begin{tikzpicture}[->,>=stealth', auto, node distance=1.75cm, semithick]
\tikzstyle{every state}=[text=black]
	\node[state](X) {$X$};
	\node[state](M) [below left of=X]{$M$};
	\node[state](A) [below right of=X]{$A$};
	\node[state](E) [below of=A]{$E$};
		
	\path[] (X) edge         node{}(M);
	\path[] (X) edge         node{}(A);
	\path[] (A) edge         node{}(E);
\end{tikzpicture}}
\caption{The one-source system with universality.} \label{fig3}
\end{figure}
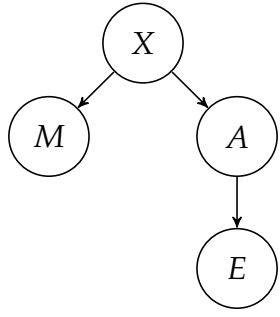

Under the conditions of our assumptions, the simplified Bayesian network given in Figure~\ref{fig3} will then adequately model the chain of inferences involved in analogue simulation supported by universality arguments. We would like to show that $\t E$ confirms $\t M$ within a Bayesian theory of confirmation. This requires that one proves that $P(\t M|\t E) > P(\t M)$. For this purpose we need to specify the prior probability of the `parent node' in the Bayesian network, i.e.
\begin{equation} \label{priorx}
P(\t X)=: x,
\end{equation}
and the conditional probabilities of all `child nodes', given the values of their parents, i.e.
\begin{eqnarray} \label{priormae}
P(\t M|\t X)=: m_x &,& P({\rm M |  \neg X})=: m_{\bar{x}} \nonumber\\
P(\t A|\t X)=: a_x &,& P({\rm A |  \neg X})=: a_{\bar{x}}\\
P(\t E|\t A)=: e_a &,& P({\rm E |  \neg A})=: e_{\bar{a}} \, .\nonumber
\end{eqnarray}

The first central assumption is that the prior probability of $\t X$ lies in the open interval $(0,1)$. However, as we discussed in Sect. \ref{unschutz}, we have theoretical arguments, in particular the result by Unruh and Sch\"{u}tzhold, in favour of $\t X$. So a rational agent would assign
\begin{equation}
1/2 < x < 1 \label{x}.
\end{equation}
The conditional probabilities are then constrained by the following conditions:
\begin{eqnarray}
m_x &>& m_{\bar{x}} \label{m}\\
a_x &>& a_{\bar{x}}\label{a}\\
e_a &>& e_{\bar{a}} \label{e}
\end{eqnarray}
The conditions (\ref{x}) to (\ref{a}) encode probabilistically the relevant elements of the universality argument since they allow for the possibility of a background assumption that supports both M and A. The statement (\ref{e}) encodes probabilistically that the empirical evidence actually plays the role of evidence in favour of the modelling framework ${\cal A}$.    

With this, the following theorem holds (the proof is in Appendix \ref{app1}):
\begin{theorem} \label{thm1}
Consider the Bayesian network depicted in Figure \ref{fig3} with the prior probability distribution $P$ defined in eqs. (\ref{priorx}) and (\ref{priormae}). Then $P(\t M | \t E)>P(\t M)$, if conditions (\ref{x}) to (\ref{e}) are satisfied. 
\end{theorem}
Theorem \ref{thm1} implies that $\t E$ confirms $\t M$ within a Bayesian analysis of confirmation. Note that Theorem \ref{thm1} also holds if condition (\ref{x}) is replaced by the weaker condition $0 < x < 1$.  Hence, within the framework of analogue simulation, provided we have universality arguments with prior probability that is neither unity or zero, confirmation of a hypothesis regarding the target system can obtain based upon evidence relating to the source system. 

It is important to note again that having independent grounds on which to support one of the common background assumptions will `block' the inductive support E can give for M as that background assumption already has a large marginal probability. This does not pose a problem for this account but offers a way to distinguish between  those circumstances in which the novel empirical evidence E can provide substantial inductive support for M and those circumstances it cannot  be used for that purpose. 

An important implication of the Bayesian analysis relates to the role of the syntactic isomorphism. The structure of the Bayesian network is such that the syntactic isomorphism is not explicitly represented. Furthermore, based upon the network, even if no syntactic isomorphism obtains between the modelling frameworks $\mathcal{M}$ and $\mathcal{A}$, one could sensibly talk about confirmation of $\t M$ by $\t E$, provided there exists some non-empty set of shared  background assumptions. The key point is that in such circumstances although confirmation of $\t M$ would indeed obtain, there would be no `analogue simulation'. As discussed above, the role of the isomorphism is to guarantee that there will be a term within the modelling language of $\mathcal{M}$ that is counterpart to the term within $\mathcal{A}$ that refers to the empirical evidence specified in variable $E$. Without such a term within $\mathcal{M}$ there would be no sense in which $\mathcal{S}$ is acting as a simulator for the behaviour of $\mathcal{T}$. Although the syntactic isomorphism is not explicitly represented in the network, it is implicitly stipulated within the universality argument. Furthermore, as noted above, the universality argument is exactly the reason we take the syntactic isomorphism to be indicative of a deeper fact about nature. Although it cannot be used to establish confirmation, the syntactic isomorphism is a crucial, if ultimately non-fundamental, heuristic for finding the universality argument that can. 

To recapitulate, in this section we have demonstrated that confirmation via analogue simulation obtains within a  Bayesian analysis provided there exists an inferential connection between the conditions of applicability of the target and system models. That is, if there exists a binary variable that is assumed to be positively correlated with the empirical adequacy of both the source and target models, then evidence in favour of the model of the source system can be used to make inferences about the target system. This, in-and-of-itself, is not a particularly surprising result, and certainly the demonstration of such in principle inferential relations is not a validation of the framework for analogue simulation that is being proposed. Rather, we take the results of this section to: i) demonstrate the internal consistency of the informal arguments towards confirmation via analogue simulation given in \shortcite{dardashti2015confirmation}; and ii) provide a powerful evaluative and heuristic tool for the analysis of analogue simulation as it exists within contemporary scientific practice. Two natural directions of further development are: i) the identification and evaluation of potential cases of confirmation via analogue simulation in other scientific examples; and ii) the refinement of the Bayesian model to include cases within more than one analogue system. The second of these will be pursued in the following section. 

\subsection{Multiple Source Confirmation}

The systems $\mathcal{S}$ and $\mathcal{T}$ discussed in the previous section were assumed to be different \textit{types} of systems and thus have fundamentally different material constitution. In the context of analogue experiments for Hawking radiation we could think of $\mathcal{S}$ as a BEC analogue back hole and $\mathcal{T}$ as an astrophysical black hole.  Of course, as already noted, there are in fact a variety of further types of analogue black holes in addition to the BEC. For example, rather than making use of the syntactic isomorphism between BEC and gravitational models we might draw inferences based upon analogue black holes constructed out of traveling refractive index interfaces in nonlinear optical media or `slow light' in moving media \shortcite{carusotto:2008}. 

With such examples in mind, we can extend the analysis of the previous section to consider the case when we have multiple types of sources each providing independent evidence for the target system modelling framework. The expectation would be that adding more source systems should increase the degree of confirmation, but that this increase will eventually reach some `saturation point' when we cannot learn any more about the target system by conducting additional analogue experiments. 

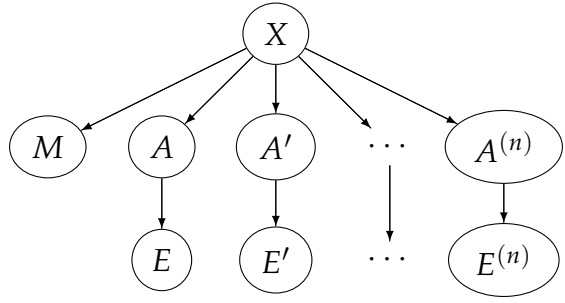
\begin{figure}
\centering
\begin{tikzpicture}
[mynode/.style={draw,ellipse,text width=2cm,-latex,text centered}]
  \node [ellipse,draw] (X) {$ X$}
  child {node [ellipse,draw] (M) {$ M$}}
   child {node [ellipse,draw] (A) {$ A$}
   	child {node [ellipse,draw] (E) {$ E$}}}
   child {node [ellipse,draw] (A1) {$ A'$}
   	child {node [ellipse,draw] (E1) {$ E'$}}}
   child {node (A') {$\cdots$}
   	child {node  (E') {$\cdots$}}}
    child {node [ellipse,draw] (An) {$ A^{(n)}$}
   	child {node [ellipse,draw] (En) {$ E^{(n)}$}}};
\path (X) edge[-latex] (M)
(X) edge[-latex] (A) 
(A) edge[-latex] (E)
(X) edge[-latex] (A1)
(A1) edge[-latex] (E1)
(X) edge[-latex] (A')
(A') edge[-latex] (E')
(X) edge[-latex] (An)
(An) edge[-latex] (En)
;
\end{tikzpicture}
\caption{The $n$-source system with universality.} \label{fig4}
\end{figure}

Consider a Bayesian network for an $n$-source system (Figure~\ref{fig4}). The question we would like to answer is how does the confirmation measure change as one increases the number of different analogue systems providing us with evidence. 
Following the same line of reasoning as the last section we assume:
\begin{eqnarray}
a'_x > a'_{\bar{x}}  \label{a'} &,& e'_{a'} > e'_{\bar{a}'}  \nonumber \\
a''_x > a''_{\bar{x}}  \label{a''} &,& e''_{a''} > e''_{\bar{a}''}  \label{ep}\\
 & \vdots &  \nn \\
a^{(n)}_x > a^{(n)}_{\bar{x}}   &,& e^{(n)}_{a^{(n)}} > e^{(n)}_{\bar{a}^{(n)}}  \nonumber
\end{eqnarray}

We can now calculate the corresponding difference measure of confirmation, which is defined as
\begin{equation}
\Delta^{(n)} := P(\t M|\t E,\t E',\dots,\t E^{(n)}) - P(\t M). 
\end{equation}
It satisfies the following theorem (the proof is in Appendix \ref{app2}):

%
\begin{theorem} \label{thm2}
Consider the Bayesian Network depicted in Figure \ref{fig4} with the prior probability distribution $P$ defined analogously to Theorem \ref{thm1}. Then $\Delta^{(n)} > 0$ is a strictly increasing function of the number of source systems, if conditions (\ref{x}), (\ref{m}), and (\ref{ep}) are satisfied. 
\end{theorem}
This theorem implies that as the number of different analogue systems providing evidence increases so does the degree of confirmation.\footnote{Note that also Theorem \ref{thm2} holds if condition (\ref{x}) is replaced by the weaker condition $0 < x < 1$.}
Again, this is not a particularly surprising result. Given that confirmation via analogue simulation obtains for a single source system, one would expect that adding in more and more (independent) source systems would allow one to increase the degree of confirmation.  The feature that is most interesting is not the fact that $\Delta^{(n)}$ is strictly increasing, but rather the functional form of this increase. In particular, the natural intuition is that as the number of source systems increases the  increase in the degree of confirmation would eventually saturate.  One of the chief virtues of the Bayesian model for analogue simulation with multiple source systems is that it allows us to give an analytical expression for such a saturation point. 

First, let us consider how $\Delta^{(n)}$ changes in the large $n$ limit. A little analytical work (again see Appendix \ref{app2}) allows us to show that\footnote{Here (and below) we use the convenient notation $\overline{x} := 1-x$.}
\begin{equation}
\lim_{n \to \infty} \Delta^{(n)} = \overline{x} \, (m_x - m_{\bar{x}}) =: \Delta_{sat} \, .
\end{equation}
This means that the maximum amount of confirmation one can obtain by adding in more and more sources is bounded by some finite value, viz. $\Delta_{sat}$, determined by the prior probabilities $x$, $m_x$ and $m_{\bar{x}}$. Beyond this point, there is vanishingly small added value (in terms of confirmation) achieved by adding in more source systems. Two features of $\Delta_{sat}$ are worth remarking on. First, the higher the prior probability of X the lower the saturation point will be. This makes sense because the more sure we are of X to start with, the lower the limit on the extra information we can learn from $\t E,\t E',\dots,\t E^{(n)}$. Second, the higher the relative likelihood of $\t M$ given $\t X$ to $\t M$ given $\neg \t X$ (i.e. $m_x-m_{\bar{x}}$), the higher the saturation point. This makes sense because the stronger the relationship between $X$ and $M$ the more we can potentially learn from $\t E,\t E',\dots,\t E^{(n)}$.

\begin{figure}
\centerline{
\includegraphics[height=8cm]{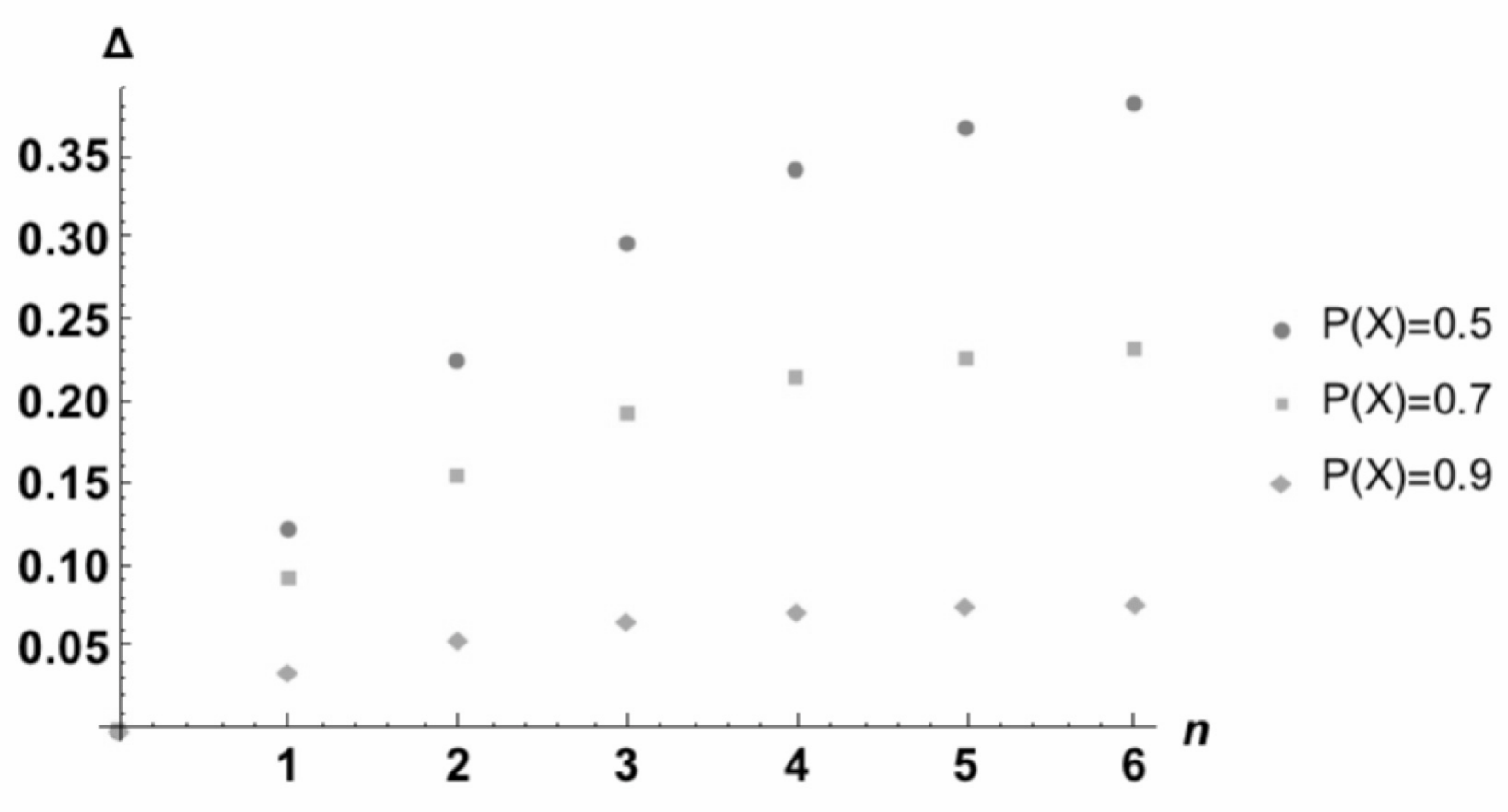}}
\caption{
The confirmation measure $\Delta^{(n)}$ as a function of $n$  for  \mbox{$m_x - m_{\bar{x}} = 0.8$} and \mbox{$\gamma_n  = 1.9^n$} and   three different priors of X.}\label{fig5}
\label{ndep}
\end{figure}

A further interesting feature that we can examine is the speed with which the saturation point is approached. We can examine this `rate of saturation' by plotting $\Delta^{(n)}$ (as specified in eq. (\ref{deltan}) in Appendix \ref{app2}) for a set of  prior probabilities of $\t X$. As can be seen from Figure~\ref{ndep}, the higher the prior probability of $\t X$, the quicker the saturation point is reached. Strikingly, for the values of the parameters considered, we find that given a prior of greater than $0.5$ for $\t X$, saturation can be reached after only three or four successful analogue experiments. 

This result is in tune with scientific intuitions regarding analogue simulation in the context of universality arguments. Consider, in particular, the dumb hole Hawking radiation case. Let us assume that an uncontested and repeated experiment demonstrating BEC Hawking radiation has already been carried out. Given initial confidence in the universality arguments, another different implementation of a source system displaying the Hawking effect (say in nonlinear optics)  would strongly increase the belief in the astrophysical Hawking effect. However, once a few such examples were constructed, one would quickly stop gaining new insight. Conversely, given initial skepticism regarding the universality arguments, a second implementation of the dumb hole source system would not  radically increase the belief in the astrophysical Hawking effect. Furthermore, in such circumstances it would only be after a diverse and extensive range of implementations of source systems that one would stop believing that new examples gave new information. Significantly, the saturation in confirmation  indicates that, under plausible assignments of priors, even an extraordinarily large range of diverse analogue experiments will not lead to conclusive confirmation of astrophysical Hawking radiation. This is in tune with the scientific intuitions that there is a limit to what can be learned about astrophysical Hawking radiation via analogue experiments.   

\subsection{Confirmation Dependence on Source System}

Although most of the proposed analogue models of black hole Hawking radiation have not been tested yet, there can be significant differences in our prior belief regarding the adequacy of these models. One reason for this is that the modelling may rely on a strong theoretical basis in one system but a rather conjectural basis in the other. For example, the fundamental physics of BECs is better understood, via the Bogoliubov theory, than that of superfluid helium-3. Another reason is that one may have good control of the experimental setup, such that one has more reason to rely on the adequate realization of the various idealizing assumptions involved in the derivation of the model. 

Such differences imply that we may plausibly assign significantly different marginal probabilities to the analogue models. In such circumstances, it is then interesting to consider the degree to which variations in the marginal probabilities one assigns to the adequacy of the analogue model may affect the confirmation behaviour of the analogue setup. The following theorem addresses this question for the one-source model (see Appendix \ref{app3} for details and the proof):
%
\begin{theorem} \label{thm3}
Consider the Bayesian Network depicted in Figure \ref{fig3} with the prior probability distribution $P$ defined in eqs. (\ref{priorx}) and (\ref{priormae}). Let $\Delta := P(\t M | \t E) - P(\t M)$ and $a :=P({\rm A})$. Then $d\Delta /d a < 0$, if conditions (\ref{x}) to (\ref{e}) hold.
\end{theorem}

This plausible result implies that an assignment of a  higher probability to the adequacy of the analogue model will have the effect of a decrease in the confirmation of the adequacy of the target model by the observation of the analogue  Hawking effect. Or to put it differently: the more certain we are about the adequacy of the analogue model that describes the source system we are experimenting on, the less effective is the evidence obtained there in confirming the adequacy of the model of the target system. Significantly, this result has direct implications for the respective epistemic value of future analogue experiments. In particular, all else being equal, it implies that scientists will learn more about the target system by conducting future analogue experiments using media about which we are \textit{less certain} regarding their fundamental physics (i.e. marginal of $A$ is lower), than those using media about which we are more confident regarding their fundamental physics (i.e. marginal of $A$ is higher).  



\section{Conclusion and Prospectus}

History is replete with examples of `transformative' technology having a profound and lasting impact on the methodological foundations of science. Much recent literature in the philosophy of science has focused on the sense in which computer simulation should be taken to have had such an impact.\footnote{See for example \cite{hartmann:1996,humphreys:1995,humphreys:2004,parker:2009,winsberg:1999,winsberg:2010}.} Analogue simulation is a new inferential tool found at the cutting edge of modern science that we see good reasons to take as potentially transformative. Building upon the initial analysis of \shortcite{dardashti2015confirmation}, in this paper we have applied a Bayesian analysis to explicate the structure of inferences that analogue simulation can and cannot allow us to make. 

Our three principal results are: i) that `single source' confirmation via analogue simulation can obtain under certain conditions; ii) that `multiple source' confirmation via analogue simulation displays the generic feature of saturation in confirmatory power; and iii) that analogue experiments in which we are more confident about the fundamental physics provide less confirmation via analogue simulation than those about which we are more confident. Our results provide a basis to both formalise the epistemic value of analogue experiments that have been conducted and to advise scientists as to the respective epistemic value of future analogue experiments.


\begin{appendix} 

\section{Proofs}
\subsection{Theorem \ref{thm1}}\label{app1}

We have to show that 
\begin{equation}
P(\t M|\t E)=\frac{P(\t M,\t E)}{P(\t E)}>P(\t M). 
\end{equation}
To do so, we consider the Bayesian network depicted in Figure~\ref{fig2} and follow the general methodology described in \shortcite[Sect. 3.5]{bovens2003bayesian}. First, we calculate
\begin{eqnarray}
P(\t M,\t E)&=&\sum_{A, X} \, P( X, \t M,  A, \t E) \nn \\ 
 &=& \sum_{A, X} \, P( X) \, P(\t M|  X) \, P( A| X) \, P(\t E| A) \nn \\ 
 &=&  x \, m_x \,  (a_x e_a + \bar{a}_x  \, e_{\bar{a}}) + \overline{x} m_{\bar{x}} \, ( a_{\bar{x}}  \, e_a +  \bar{a}_{\bar{x}}  \, e_{\bar{a}}) \nn \\
 &=& x \, m_x  \,  \alpha + \overline{x}  \, m_{\bar{x}} \,  \beta,
\end{eqnarray}
with
\begin{equation}  \label{alpha} 
\alpha :=  a_x  \, e_a + \bar{a}_x  \, e_{\bar{a}}   \quad , \quad  \beta :=   a_{\bar{x}}  \, e_a +  \bar{a}_{\bar{x}}  \, e_{\bar{a}}  .  
\end{equation}
Note that $\alpha = P({\rm E | X})$ and $\beta = P({\rm E | \neg X})$. Similarly we obtain
\begin{eqnarray}
P(\t E) &=& \sum_{A,  M, X} P( X,  M,  A, \t  E)\nn \\ 
&=&     x  \,  \alpha + \overline{x}  \, \beta
\end{eqnarray}
and
\begin{eqnarray}
P(\t M) &=& \sum_{A,  E, X} P( X,  \t M,  A, E)\nn \\ 
&=& x \, m_x + \overline{x} \, m_{\bar{x}}.
\end{eqnarray}

Next, we define the difference measure $\Delta:=P(\t M|\t E) - P(\t M)$ and obtain after some algebra:
\begin{eqnarray}
 \Delta &=& \frac{ x  \, m_x \,  \alpha + \overline{x}  \, m_{\bar{x}} \, \beta -  (x  \, m_x + \overline{x}  \, m_{\bar{x}}) \, (x  \,  \alpha + \overline{x}  \, \beta)}{  x   \, \alpha + \overline{x} \,  \beta }\nn \\
&=& \frac{x \, \overline{x}}{  x  \,  \alpha + \overline{x}  \, \beta }  \cdot  (m_x - m_{\bar{x}}) \, (\alpha -\beta)
\end{eqnarray}
With
\begin{equation}
\alpha -\beta = (a_x - a_{\bar{x}}) \, (e_a - e_{\bar{a}})   \label{ab}
\end{equation}
it follows that
\begin{equation} \label{delta2}
\Delta = \frac{x \, \overline{x}}{x \,  \alpha + \overline{x}  \, \beta } \cdot (a_x - a_{\bar{x}}) \, (e_a - e_{\bar{a}}) \, (m_x - m_{\bar{x}}) .
\end{equation}

Hence, $\Delta>0$, if conditions (\ref{x}) to (\ref{e}) are satisfied. This completes the proof of Theorem \ref{thm1}.

\subsection{Theorem \ref{thm2}} \label{app2}

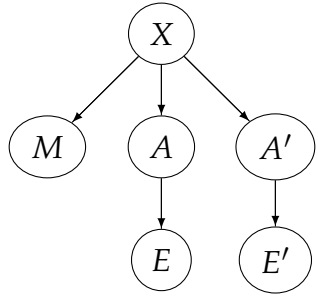
\begin{figure}
\centering
\begin{tikzpicture}
[mynode/.style={draw,ellipse,text width=2cm,-latex,text centered}]
  \node [ellipse,draw] (X) {$X$}
  child {node [ellipse,draw] (M) {$M$}}
   child {node [ellipse,draw] (A) {$A$}
   	child {node [ellipse,draw] (E) {$E$}}}
    child {node [ellipse,draw] (A1) {$A'$}
   	child {node [ellipse,draw] (E1) {$E'$}}};
\path (X) edge[-latex] (M)
(X) edge[-latex] (A) 
(A) edge[-latex] (E)
(X) edge[-latex] (A1)
(A1) edge[-latex] (E1)
;
\end{tikzpicture}
\caption{The 2-source system with universality.} \label{2source}
\end{figure}

To see how Theorem \ref{thm1} can be generalized to the $n$-source systems represented in Figure~\ref{fig4} let us consider first the simplified 2-source system represented in Figure~\ref{2source}. Here we need to show that $P(\t M|\t E,\t E')= P(\t M,\t E,\t E') / P(\t E,\t E') > P(\t M)$. To do so, we first calculate 

\begin{eqnarray}
P(\t M,\t E,\t E')&=&\sum_{A, A', X} P( X, \t M,  A, \t E, A', \t E') \nn \\ 
&=&  \sum_{A, A', X} \, P(X) \, P(\t M| X) \, P(A| \t X) \, P(\t E| A) \, P(A'| X) \, P(\t E'| A') \nn \\ 
&=&  x  \, m_x  \, \alpha \,  \alpha'   +   \overline{x}  \, m_{\bar{x}}  \,  \beta \,  \beta' ,
\end{eqnarray}
where $\alpha'$ and $\beta'$ are  defined analogously to eqs. (\ref{alpha}) with $e$ and $a$ replaced by $e'$ and $a'$.

Similarly we obtain
\begin{eqnarray}
P(\t E,\t E') &=& \sum_{ A, A', X}  \, P( X) \, P(\t E| A) \, P( A| X) \,  P(\t E'| A') \, P( A'| X)          \nn \\ 
&=&     x \,   \alpha \,  \alpha' + \overline{x} \,  \beta  \, \beta'.
\end{eqnarray}
Defining $\Delta^{(2)} := P(\t M|\t E,\t E')-P(\t M)$ it follows that
\begin{equation} \label{delta'} 
\Delta^{(2)} = \frac{x \, \overline{x}  }{x  \,  \alpha \, \alpha' + \overline{x}  \, \beta \,  \beta'}  \cdot (m_x - m_{\bar{x}}) \, (\alpha  \, \alpha' -\beta  \, \beta')     .
\end{equation}
Now it is easy to see that $\alpha \,  \alpha' -\beta \,  \beta' > 0$ iff $(\alpha / \beta) \cdot  (\alpha'  / \beta')>1$. This holds if $\alpha > \beta$ and $\alpha' > \beta'$.  Both conditions hold because of assumptions (\ref{ep}). Hence, $\Delta^{(2)} > 0$, if conditions (\ref{x}), (\ref{m}), and (\ref{ep}) are satisfied. 

It is straightforward to generalise this calculation to the $n$-source system represented in Figure~\ref{fig4}. We obtain
\begin{eqnarray}\label{deltan}
\Delta^{(n)} &=& \frac{ x  \, \overline{x} \, (m_x - m_{\bar{x}}) \,  \left(\prod_{k=1}^n  \alpha^{(k)} -   \prod_{k=1}^n \beta^{(k)} \right) }{x \prod_{k=1}^n \alpha^{(k)}   + \overline{x}  \,  \prod_{k=1}^n   \beta^{(k)} } \nonumber \\
&=& \frac{ x  \, \overline{x} }{x \gamma_n  + \overline{x} } \cdot (m_x - m_{\bar{x}}) \cdot \left(\gamma_n -  1 \right), \label{confmeasure}
\end{eqnarray}
with the likelihood ratio $\gamma_n :=\prod_{k=1}^n \alpha^{(k)}/\beta^{(k)}$. Note that $\alpha^{(k)} > \beta^{(k)}$ for all $k$ implies that $\gamma_n > 1$. Hence, as before, $\Delta^{(n)} >0$, if conditions (\ref{x}), (\ref{m}), and (\ref{ep}) are satisfied. 

Note that $\gamma_n$ strictly monotonically increases with $n$. Furthermore,
\begin{equation}
\frac{\partial \Delta^{(n)}}{\partial \gamma_n} =  \frac{x  \, \overline{x}  \,  (m_x - m_{\bar{x}}) }{\left(x  \, \gamma_n + \overline{x}\right)^2} >0,
\end{equation}
if conditions (\ref{x}), (\ref{m}), and (\ref{ep}) are satisfied. Hence, $\Delta^{(n)}$ is a positive function of $n$. This completes the proof of Theorem  \ref{thm2}.

In closing this section, we investigate the limiting behaviour of $ \Delta^{(n)}$ as a function of $n$. Setting \mbox{$\kappa :=x\, \overline{x} \, (m_x - m_{\bar{x}})$}, we obtain
\begin{eqnarray}
\lim_{n \to \infty} \Delta^{(n)} &=& \lim_{x \to \infty} \kappa \, \frac{\gamma_n -1}{x \,  \gamma_n + \overline{x}} \nonumber \\
&=&  \lim_{x \to \infty} \kappa \, \frac{1-1/\gamma_n}{x   + \overline{x}/\gamma_n} \nonumber  \\
&=& \kappa/x  \nonumber  \\
&=&\overline{x} \, (m_x - m_{\bar{x}}),
\end{eqnarray}
where we have used the fact that $\lim_{n \to \infty}  \gamma_n = \infty$.

\subsection{Theorem  \ref{thm3}}\label{app3}

The proof follows from a straight-forward computation starting from eq. (\ref{delta2}):
\begin{eqnarray}
\frac{d\Delta}{d a} &=&      
			  \frac{\partial \Delta}{\partial a_x} \cdot \frac{d a_x}{d a} 
			+   \frac{\partial \Delta}{\partial a_{\bar{x}}} \cdot  \frac{d a_{\bar{x}}}{d a}  
			+ \frac{\partial \Delta}{\partial x} \cdot  \frac{d x}{d a}  \,   \nonumber\\
			&=&
			\frac{1}{x} \cdot  \frac{\partial \Delta}{\partial a_x} 
			+ \frac{1}{\overline{x}}  \cdot \frac{\partial \Delta}{\partial a_{\bar{x}}} 
			+ \frac{1}{(a_x - a_{\bar{x}})}  \cdot \frac{\partial \Delta}{\partial x}, \label{eq1}
\end{eqnarray}
where we have used that $a := x  \, a_x + \overline{x}  \, a_{\bar{x}}$. Taking the respective partial derivatives yields
\begin{eqnarray*}
 \frac{\partial \Delta}{\partial a_x} &=&      \frac{k  \, x  \, \overline{x}}{(x \,  \alpha + \overline{x} \,  \beta )^2} \cdot \bigg(x \,  \alpha + \overline{x} \,  \beta - (a_x - a_{\bar{x}}) (e_a - e_{\bar{a}})  \, x \bigg)  \\
  \frac{\partial \Delta}{\partial a_{\bar{x}}} &=&     -  \frac{k  \, x \, \overline{x}}{(x  \, \alpha + \overline{x}  \, \beta )^2}  \cdot  \bigg(x  \, \alpha + \overline{x} \,  \beta - (a_x - a_{\bar{x}}) (e_a - e_{\bar{a}})  \, \overline{x} \bigg)   \\
  \frac{\partial \Delta}{\partial x} &=&  -  \frac{k  \, (a_x - a_{\bar{x}}) }{(x \,  \alpha + \overline{x}  \, \beta )^2} \cdot 
  \bigg(x^2 \,  \alpha - \overline{x}^2 \, \beta  \bigg).
\end{eqnarray*}
Here we have used the shorthand $k := (e_a - e_{\bar{a}}) \, (m_x - m_{\bar{x}})$. Plugging these expressions into eq. (\ref{eq1}) one obtains after some algebra that
\begin{equation}
\frac{d \Delta}{d a} =  - \frac{k }{(x  \, \alpha + \overline{x} \,  \beta )^2} \cdot \bigg(x \, (3 \, x -1)  \, \alpha -  \overline{x} \, (3 \, \overline{x} -1)  \, \beta \bigg). \label{vari}
\end{equation}
Assumptions (\ref{m}) and (\ref{e}) imply that $k > 0$ and assumptions (\ref{x}), (\ref{a}) and (\ref{e}) imply that the large bracket is positive. (To see this, note that \mbox{$x \, (3 \, x -1) >  \overline{x} \, (3 \, \overline{x} -1)$} for \mbox{$1/2 < x < 1$}.) Hence, $d \Delta/d a < 0$. This completes the proof of Theorem  \ref{thm3}.


 \end{appendix}

\bibliographystyle{chicago}
\bibliography{dumb}
\end{document}